\documentclass[useAMS,usenatbib,usegraphicx]{mn2e}
\title[Non-uniform stellar distribution in clusters] {Winds from clusters with non-uniform stellar distributions}
\author[Rodr\'iguez-Gonz\'alez, A. et al.]{Rodr\'iguez-Gonz\'alez,
A.$^1$ \thanks{ary@nucleares.unam.mx}; Cant\'o, J.$^2$;
Esquivel, A.$^1$,  Raga, A. C.$^1$ \& Velazquez, P. F.$^1$\\
$^1$Instituto de Ciencias Nucleares, Universidad
Nacional Aut\'onoma de M\'exico, Apdo. Postal 70-543,
04510, M\'exico, D.F., M\'exico.\\
$^2$Instituto de Astronom\'\i a, Universidad Nacional
Aut\'onoma de M\'exico, Apdo. Postal 70-264, 04510, M\'exico, D.F., M\'exico.}

\begin{document}
\date{Draft Version, \today}
\pagerange{\pageref{firstpage}--\pageref{lastpage}} \pubyear{2007}
\maketitle
\label{firstpage}
\begin{abstract}
We present analytic and numerical models of the `cluster wind'
resulting from the multiple interactions of the winds ejected
by the stars of a dense cluster of massive stars. We consider the
case in which the distribution of stars (i.~e., the number of stars
per unit volume) within the cluster is spherically symmetric,
has a power-law radial dependence, and drops discontinuously
to zero at the outer radius of the cluster. We carry out comparisons
between an analytic model (in which the stars are considered in terms
of a spatially continuous injection of mass and energy) and 3D
gasdynamic simulations (in which we include 100 stars with identical
winds, located in 3D space by statistically sampling the stellar
distribution function). From the analytic model,
we find that for stellar distributions
with steep enough radial dependencies the cluster wind flow develops
a very high central density and a non-zero central velocity, and
for steeper dependencies it becomes fully supersonic throughout
the volume of the cluster (these properties are partially reproduced
by the 3D numerical simulations). Therefore, the wind solutions
obtained for stratified clusters can differ dramatically from the
case of a homogeneous stellar distribution (which produces
a cluster wind with zero central velocity, and a fully subsonic flow
within the cluster radius). Finally, from our numerical simulations we
compute predictions of X-ray emission maps and luminosities, which can
be directly compared with observations of cluster wind flows.
\end{abstract}

\begin{keywords}
Hydrodynamics -- shock waves -- stars: winds, outflows
\end{keywords}
\section{Introduction}
Super star clusters (SSCs) are dense clusters of young massive stars,
 first identified in NGC 1705 by Melnick et al. (1985) and in NGC 1569 by Arp \& Sandage (1985). Recently,
they have been observed in a wide range of star-forming galaxies, such as
merging systems  (NGC 4038/4039, Whitmore \& Schweizer 1995),
dwarf galaxies (Henize 2-10, Johnson et al. 2000), classical
starbursts (M82, Gallagher \& Smith 1999, Melo et al. 2005),
as well as in our galaxy (Arches Cluster~: Nagata et al. 1995; Cotera
et al. 1996; Serabyn, Shupe \& Figer 1998) amongst many other
systems (for a review see Whitmore 2000).

These star clusters can contain hundreds or thousands of very young,
energetic stars, and have stellar densities far greater than those
seen in normal OB associations. The ages of most of these star clusters are
around 1- 10 Myr, their radii typically in the range of $\sim$ 1 -10 pc,
and their total cluster masses in the $10^3$ - $10^6$ M$_\odot$ range (Melo et al. 2005 reported a  mean 
mass per star cluster, of $\sim 2 \times 10^5$ $M_\odot$, for M82).
The central stellar densities of SSCs reach up to $\sim$ $10^5$ M$_\odot$
pc$^{-3}$. However, we can find SSCs with older ages and/or larger masses (Walcher et al. 2006 reports 
a cluster with 6 $\times$ 10$^7$ $M_\odot$).

Cant\'o et al. (2000, hereafter Paper I) explored an analytic model
using the physical properties of the {\it Arches Cluster}. This cluster
has $\sim 100$ massive stars within a $\sim$ 0.2 - 0.3 pc radius.
Such SSC must have strong multiple stellar wind interactions,
resulting in the formation of stellar wind shocks. In Paper I, the cluster wind
was modeled as a mass-loading process (Hartquist et al. 1986;
Dyson 1992; Lizano et al. 1996), resulting in essentially the same
model as the one presented by Chevalier \& Clegg (1985, who studied
the flow resulting from a continuous temporal and spatial distribution
of supernovae explosions).

Both in the stationary solution for spherically symmetric winds (Chevalier
\& Clegg 1985, Cant\'o et al. 2000) and in the numerical calculations
of Raga et al. (2001) the stellar distribution (within the cluster)
was assumed to be homogeneous. Also, these models are adiabatic (or,
more precisely, non-radiative) solutions, which are appropiate for SSCs with low
to intermediate mass and/or terminal velocity of SSCs
(masses around $10^4$ - $10^6$ $M_\odot$ and terminal velocities of
$\sim 1000$ km/s). For more massive stellar clusters, lower stellar wind
terminal velocities or higher metallicities, radiative losses within the
cluster wind may become important (see Silich et al. 2004).

In the present paper, we explore a non-radiative analytic model for
cluster winds, considering a non-homogeneous stellar distribution
(within the cluster). In particular, we study the case in which
the stars have a $n(R)\propto R^\alpha$ power-law distribution
(where $n$ is the number of stars per unit volume as a function of the
spherical radius $R$), with $-3<\alpha\leq 0$. We also compute 3D
gasdynamical simulations for stellar distribution functions with
different values of $\alpha$, and compare the properties of the computed
flows with the analytic model.
This work is a natural extension of the model presented in Paper I.

We note that Matvienko \& Shchekinov (2005) have presented a
study of mass loaded flows with sources and sinks with power
law spatial distributions. The analytic model that we discuss
in the present paper is based on a similar set of equations,
but describes a cluster wind, rather than a mass loaded stellar
wind (as was studied by Matvienko \& Shchekinov 2005). Also relevant
in the context of the present work are the papers of Raga et al. (2001)
and Rockefeller et al. (2005), who carried out 3D numerical simulations
of winds from stratified clusters.

The paper is organized as follows. In section 2, we present the
analytical solution. In section 3, we describe the numerical simulations
and compare the results with the analytic model. Predictions of
the X-ray emission from the simulated flows are described
in section 4. Finally, we summarize
our results in section 5.\\

\section{The analytic model}\label{sec:analytic}

We consider $N$ identical stars in a spherical
cluster with an outer radius $R_c$. The stars have a spatial
distribution (number of stars per unit volume) of the form,

\begin{equation}
\label{eq:StellarDens}
n(R)=k_c R^\alpha=\frac{(3+\alpha)N}{4\pi R^{3+\alpha}_c} R^\alpha \,,
\end{equation}

where $R$ is the spherical radius, $\alpha$ and $k_c$ are constants.
In the second equality, the constant $k_c$ has been
computed using the normalization condition

\begin{equation}
\int_0^{R_c} 4\pi R^2 n(R)\, dR=N\,.
\label{norm}
\end{equation}

Every star has an identical wind with mass and energy depositon rates
$\dot{M}_w$ and $\dot{E}_w$, respectively, and terminal velocity $V_w$. For highly supersonic wind,  $\dot{E}_w=\frac{1}{2}\dot{M}_w V^2_w.$
The stellar winds are thermalized at shocks (produced by interactions
between the multiple winds), resulting in the production of a hot
intercluster gas. This gas has a large central overpressure, which eventually produce a stationary `cluster wind' flow.\\

For an adiabatic, spherically symmetric cluster wind (and negleting the
gravity due to the stellar distribution), the mass, momentum and
energy equations are~:

\begin{equation}
\label{eq:MassConsIn}
\frac{1}{R^2} \frac{d}{dR}(\rho V R^2)=n(R) \dot{M}_w\,,
\end{equation}
\begin{equation}
\label{eq:MomConsIn}
\rho V \frac{d V}{dR}=-\frac{dP}{dR}-n(R) \dot{M}_w V\,,
\end{equation}
\begin{equation}
\label{eq:EnConsIn}
\frac{1}{R^2}\frac{d}{dR}\left[\rho V R^2
\left(\frac{V^2}{2}+h \right)\right]=n(R) \dot{E}_w\,,
\end{equation}
$\rho$ and $V$ are the mass density and velocity of the wind respectively, 
$R$ is the radial coordinate, 
\begin{equation}
\label{eq:Enthalpy}
h=\frac{\gamma}{\gamma-1} \frac{P}{\rho}\,,
\end{equation}
and $h$is the specific enthalpy,  $P$ is the mean gas pressure and
$\gamma$ is the specific heat ratio.\\

Equation~(\ref{eq:MassConsIn}) implies that
\begin{equation}
\label{eq:FluxIn}
\rho V = \frac{\dot{M_c}}{4 \pi R^{3+\alpha}_c}R^{1+\alpha}\,,
\end{equation}
\noindent
where $\dot{M}_c \equiv N \dot{M}_w$ is the total mass loss rate from the cluster in steady state. This equation shows that the mass flux profile strongly depends on the $\alpha$ exponent.\\

The adiabatic sound speed $c_s$ is given by,
\begin{equation}
\label{eq:SoundSpeed}
c^2_s=\gamma\frac{P}{\rho}=\gamma \frac{k T}{\mu},
\end{equation}
where $T$ is the gas temperature, $\mu$ is the mean mass per
particle and $k$ is the Boltzmann constant. Integrating equation~(\ref{eq:EnConsIn}) and using equations~(\ref{eq:Enthalpy}) and (\ref{eq:SoundSpeed}) we find

\begin{equation}
\label{eq:EnergySol}c^2_s=\frac{\gamma-1}{2} (V^2_w-V)\,.
\end{equation}
Now, combining equations (\ref{eq:MomConsIn}), (\ref{eq:FluxIn})
and (\ref{eq:EnergySol}) we obtain
\begin{equation}
\label{eq:DerVelIn}
\frac{1-c v^2}{a + b v^2}\frac{dv^2}{v^2}=\frac{dr^2}{r^2},
\end{equation}
where, $a=1+\alpha$, $b=(1+\alpha+[5+\alpha]\gamma)/(\gamma-1)$,
and $c=(\gamma+1)/(\gamma-1)$. In terms of the dimensionless variables, $v = V/V_w$ $r=R/R_c\,$, this equation admits solutions,
\begin{equation}
\label{eq:VelIn}
v^p\left(a + b v^2 \right)^q=A r \,\,\,\,\,\,\,\, v=(-a/b)^{1/2}=const.
\end{equation}
\noindent
In equation (\ref{eq:VelIn}), $p=1/a$, $q=-(b+a\,c)/(2a\,b)$
and $A$ is an integration constant. \\

Outside the cluster (i.~e., for $R>R_c$) $n(R)=0$, and the mass
and momentum conservation equations have the form,
\begin{equation}
\label{eq:MassConsOut}
\frac{1}{R^2}\frac{d}{dR}(\rho V R^2)=0\,,
\end{equation}
\begin{equation}
\label{eq:MomConsOut}
\rho V \frac{d V}{dR}=-\frac{dP}{dR}\,.
\end{equation}

Combining equations (\ref{eq:MassConsOut}) and (\ref{eq:MomConsOut}),
we obtain,
\begin{equation}
\label{eq:VelOut}
v \left(1-v^2 \right )^{1/(\gamma-1)} = \frac{B}{r^2},
\end{equation}
where, $B$ is a constant.\\
\noindent
The velocity of the flow, ($v_1$) at the outer boundary of the
cluster ($r=R/R_c=1$) follows from equation~(\ref{eq:VelOut}),\\
\begin{equation}
\label{eq:VelBoundaryOut}
v_1 \left(1-v^2_1 \right )^{1/(\gamma-1)} = B.
\end{equation}

The value of the integration constants $A$ and B are determined by the boundary conditions at $r=1$ ($R=R_c$) and $r \to \infty$. They, of course, depend also on the value of $\alpha$. There are three different regimes~:

\noindent {\bf a. $\alpha \geq -1$~:}\\

 In this regime, $p>0$ and $q<0$ (also both $a$ and $b>$0). From equation (\ref{eq:VelIn})
we can see that very close to the centre of the cluster,
\begin{equation}
\label{eq:VelCenter}
v \sim (A r)^{1/p}.
\end{equation}
The wind velocity inside the cluster is subsonic, the velocity at the
centre of the cluster is zero and the density is finite.
The gas velocity increases towards
the outer cluster boundary, and the flow at this boundary
($r=1$) follows from equation (~\ref{eq:VelIn}),
\begin{equation}
\label{eq:Aconst}
A=v^p_1\left[a + b v^2_1 \right]^q\,.
\end{equation}
The left-hand side of equation (\ref{eq:VelBoundaryOut}) has a maximum value of
\begin{equation}
\label{eq:MaxIn}
\left(\frac{\gamma-1}{\gamma+1} \right)^{1/2}
\left(\frac{2}{\gamma+1}\right)^{1/(\gamma-1)},
\end{equation}
for $v_1=[(\gamma-1)/(\gamma+1)]^{1/2}$. The requirement that the
pressure at infinity must go to zero implies that the flow has
to adopt the critical solution, for which
\begin{equation}
\label{eq:constA}
A=\left(\frac{\gamma-1}{\gamma+1} \right)^{p/2} \left(a+b\frac{\gamma-1}{\gamma+1} \right)^q,
\end{equation}
and,
\begin{equation}
\label{eq:constB}
B= \left (\frac{\gamma-1}{\gamma+1} \right)^{1/2} \left( \frac{2}{\gamma+1} \right)^{1/(\gamma-1)}
\end{equation}

We can then use equations~(\ref{eq:FluxIn}), (\ref{eq:SoundSpeed}),
(\ref{eq:EnergySol}) and (\ref{eq:VelCenter}) to obtain the properties
of the cluster wind at the centre of the cluster (see Paper I):
\begin{equation}
\label{eq:Den0}
\rho_0=\frac{\dot{M}_c}{4 \pi A^a R^2_c V_w}\,,
\end{equation}
%
%\begin{equation}
%\label{Sound0}
%c^2_0=\frac{\gamma-1}{2}V^2_w,
%\end{equation}
%
\begin{equation}
\label{eq:Pressure0}
P_0=\frac{\gamma-1}{2\gamma} \frac{\dot{M}_c V_w}{4 \pi A^a R^2_c}\,,
\end{equation}
\begin{equation}
\label{eq:Temp0}
T_0=\frac{\gamma-1}{2\gamma}\frac{\mu}{k} V^2_w\,.
\end{equation}

\noindent {\bf b. $\alpha_{min} \equiv - \frac{3 \gamma+1}{\gamma+1} < \alpha \leq -1$~:}\\

In this interval, $p<0$, $q>0$, $a<0$ and $b>0$. Thus the velocity inside the cluster is subsonic and has a non-zero value at the cluster centre ($r=0$). From equation~(\ref{eq:VelIn})
\begin{equation}
\label{eq:Veloc1}
v_0=\sqrt{-\frac{a}{b}}\,.
\end{equation}

On the other hand, equation~(\ref{eq:FluxIn}) indicates that the mass flux of the wind is inversely proportional to a positive power of $r$. This implies that the central gas density tend to infinite. From equations~(\ref{eq:SoundSpeed}), (\ref{eq:EnergySol}) and (\ref{eq:Veloc1}) the central temperature is,
\begin{equation}
\label{eq:Temp1}
T_0=\frac{\gamma-1}{2\gamma}\frac{\mu}{k} \left( 1+\frac{a}{b} \right) V^2_w.
\end{equation}
%\begin{equation}
%\label{Sound1}
%c^2_0=\frac{\gamma-1}{2}V^2_w \left(1+\frac{a}{b} \right)
%\end{equation}

\noindent {\bf c. $\alpha_{cr}\equiv -3 < \alpha \leq \alpha_{min}$:~}

In this regime, $p>0$, $q<0$, $a<0$ and $b>0$, the velocity of the wind remains constant (with a supersonic value) within the cluster. Its magnitud is given by equation~(\ref{eq:Veloc1}). The temperature is also uniform inside the cluster and is given by equation~(\ref{eq:Temp1}). The density also goes to infinity at the centre of the cluster and decreases outwards. But the wind escapes from the cluster surface supersonically, and
accelerates outwards, until it reaches the terminal velocity.

We do not consider power-law stellar distributions with
$\alpha<\alpha_{cr} \,(=-3)$ because they have an infinite number
of stars (resulting from the strong divergence of the distribution
function in the cluster centre).

\section{The numerical simulations}
\subsection{Numerical setup}

In order to illustrate the analytic star cluster wind solutions we have
computed 3D numerical simulations with the full, non-radiative
gasdynamic equations. The simulations solve a multiple stellar
wind interaction problem with the 3D, adaptive grid ``yguaz\'u-a''
code, which is described in detail by Raga et al. (2000, 2002). The
simulations were computed on a five-level binary adaptive grid with
a maximum resolution of 0.1172 pc (corresponding to $256^3$ grid
points at the maximum grid resolution) in a computational domain of
30 pc (along each of the 3 coordinate axes).\\

In all runs, we assumed that the computational domain was initially
filled by a homogeneous, stationary ambient medium with temperature
$T_{env}=$ 500 K and density $n_{env}=0.1$ cm$^{-3}$. The stellar winds
are imposed in spheres (centred at the stellar positions, see below)
of radius $R_w=2.2\times 10^{18}$ cm, corresponding to 6 pixels
at the maximum resolution of the adaptive grid.
Within these spheres, we impose (at all times) a $T_w=15000$~K
temperature, and an outwardly directed $V_w=1000$~km~s$^{-1}$
velocity. The density within the spheres has an $r^{-2}$ law
(where $r$ is the radial coordinate measured outwards from
the stellar position), scaled so that the mass loss rate
is ${\dot M}_w=10^{-5}$ M$_\odot$ yr$^{-1}$ for each star.
We then place 100 such stellar wind sources within a spherical
cluster of outer radius $R_c=10$~pc, centred in the computational
domain.

We have computed four simulations of clusters with power-law
stellar distributions (see equation \ref{eq:StellarDens}) with
$\alpha=0,\,-0.5,\,-2.0$ and $-2.5$. The stellar positions are
chosen by statistically sampling the distribution functions in
the way described in section \ref{sec:StarPos}.

%We have computed mean values for radial velocity, density and temperature as a function of spherical radius R, messured from the ($x,y,z$)=(0,0,0) center of the cluster. In particular the radial velocity was calculated using

%\begin{equation}
%\label{eq:RadVel}
%v_r=\frac{x v_x+ y v_y +z v_z}{R}
%\end{equation}
%where, $v_x$, $v_y$ and $v_z$ are the project velocity in $x$, $y$ and $z$ coodinates, respectively. The scalar variables (density, temperature, etc.) was calculated using the average value for radial coordinates.

\subsection{Sampling the stellar position distribution
function}\label{sec:StarPos}

In order to produce models that can be compared with the analytic solutions,
we place stars within the cluster with positions obtained by randomly
sampling the power-law distribution function given by equation
(\ref{eq:StellarDens}). This is done as follows.

We first note that $f(R)dR=C\,4\pi R^2 n(R)dR$
(with $n(R)$ given by equation \ref{eq:StellarDens})
is the fraction of the stars which
have radial positions between $R$ and $R + dR$. Using the
normalization condition $\int_0^{R_c} f(R)dR=1$, we obtain
\begin{equation}
\label{eq:Frac}
f(R)=\left({{3+\alpha}\over {R_c}^{3+\alpha}}\right)\,R^{2+\alpha}\,.
\end{equation}
This normalization is of course  valid only for $\alpha>-3$ since otherwise
a divergence occurs at the lower limit of the normalization integral.

With a random number generator we then choose a number $\eta$ which
is uniformly distributed in the interval $[0,1]$. This variable
is statistically related to the radius $R$ through the relation~:
\begin{equation}
\label{eq:IntRand}
\int^{R}_0 f(R) dR = \int^{\eta}_0 d \eta\,.
\end{equation}

From equations (\ref{eq:Frac}-\ref{eq:IntRand}) we obtain
\begin{equation}
\label{eq:RandR}
R=R_c\eta^{1/(3+\alpha)}\,
\end{equation}
from which we can sample $R$ as a function of the random number $\eta$.

Once we have chosen the radial coordinate for each of the $N$ stars (used in the simulation)
by sampling the radial distribution function (as described above),
we asign random directions to the vector position of each of
the stars. In this way, we obtain statistical samplings of the
stellar position functions.

In practice, we have to modify the obtained stellar distributions
because we impose the stellar wind conditions in spheres of a
finite radius $r_w$ (see section 2.1). Whenever we obtain pairs
of stellar positions resulting in overlapping ``stellar wind spheres'',
we elliminate one of the two stars. This leads to an undersampling
of the desired distribution function in the central, high
stellar density regions of the generated clusters. Because of
this, the comparison between the numerical simulations and the
analytic cluster wind solutions is only meaningful away from the
central region of the cluster.

We should note that the 100 stars of each model were chosen with
the same set of random numbers, so that the directions from the cluster
centre to each of the stars are the same in all models. However, the
physical radii corresponding to the ``radial random numbers'' differ
for each model, as the ``conversion'' from random number 
($\eta$) to physical radius $R$ (see equation \ref{eq:RandR}) depends
on the value of $\alpha$.

\subsection{Model results}

\begin{figure}
\centering
\includegraphics[width=65mm]{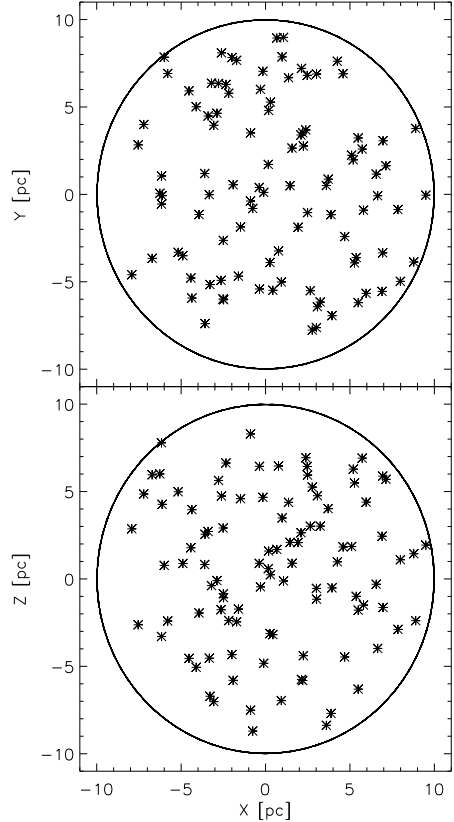}
\caption{The stellar distribution of  xy-plane and xz-plane for the
$\alpha=0$ model. The solid line represents the outer boundary
of the cluster.}\label{fig:PosR0}
\end{figure}

We computed four numerical simulations of clusters with stellar
distributions with different $\alpha$ values. All models
are computed for clusters with 100 stars and a 10 pc outer radius.
The stars are identical, with a mass deposition rate of
$10^{-5}$ M$_\odot$ yr$^{-1}$ and a wind velocity of 1000 km/s.

The only difference between the models are the stellar density
distributions. We have chosen $\alpha$ values (see equation~\ref{eq:StellarDens})
covering the different regimes of the cluster wind flow (see
section~\ref{sec:analytic}).

We have integrated forward in time the four models until stationary
flows are obtained. From the stationary flow configurations, we then
compute (through appropriate interpolations in the cartesian adaptive
grid) the radially dependent flow density, velocity and temperature
averaged over spherical concentric surfaces $S_R=4 \pi R^2$~:
\begin{equation}
\rho_a(R)={1\over 4\pi}{\int_{S_R} \rho\,\sin\theta\,d\theta d\phi}\,,
\label{ra}
\end{equation}
\begin{equation}
v_a(R)={1\over 4\pi \rho_a(R)}{\int_{S_R} \rho v_R\,
\sin\theta\,d\theta d\phi}\,,
\label{va}
\end{equation}
\begin{equation}
T_a(R)={1\over 4\pi \rho_a(R)}{\int_{S_R} \rho T\,
\sin\theta\,d\theta d\phi}\,,
\label{ta}
\end{equation}
where $\theta$ and $\phi$ are the polar and azimuthal angles, respectively, and $\rho$ is the flow density, $T$ the temperature and
$v_R$ the radial velocity (obtained by projecting the
three cartesian velocity components resulting from the numerical
integration onto the direction normal to the spherical surface). That is $v_R=(x v_x+ y v_y+ z v_z)/R$.
We then compare this spherically averaged flow with the results
from our analytic cluster wind model.
\begin{figure}
\centering
\includegraphics[width=80mm]{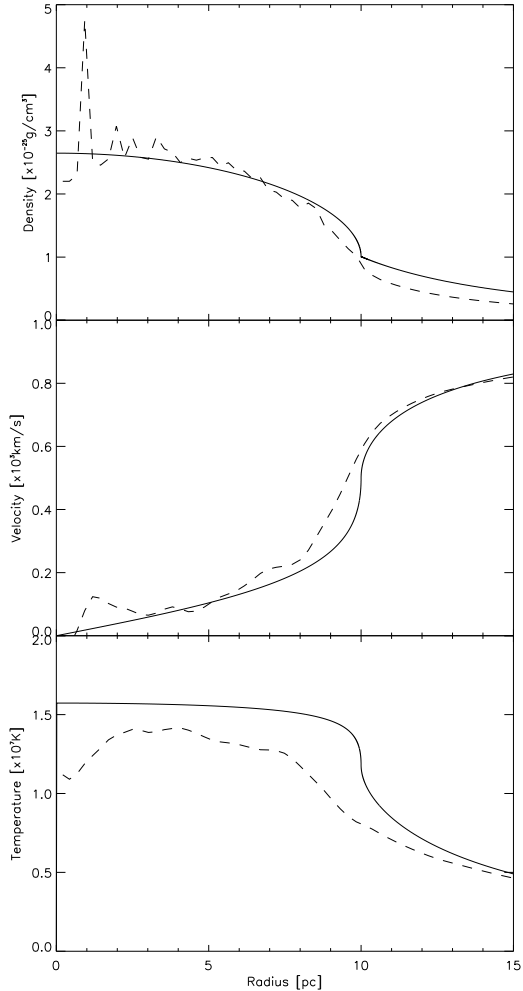}
\caption{The spherically averaged flow obtained from the $\alpha=0$ model.
The density (top), radial velocity (centre) and temperautre (bottom)
obtained from the numerical simulations (dashed lines) and the analytic
model (solid lines) are shown as a function of spherical radius $R$.}
\label{fig:R0}
\end{figure}

\subsubsection{The $\alpha \geq$ -1 case}

For this $\alpha$ range we ran two  models: a cluster with a
homogeneous (i.e., $\alpha=0$) stellar density distribution, and
a cluster with an $\alpha=-0.5$ distribution.\\

The stellar distribution used for the $\alpha=0$ model
(obtained by sampling the $\alpha=0$ distribution function,
see equation~\ref{eq:Frac} and section~\ref{sec:StarPos}) is shown in Figure~\ref{fig:PosR0}.
The average flow variables as a function of spherical radius
(obtained by averaging over concentric spherical shells, see above) are compared
with the solution obtained from the analytic model in Figure \ref{fig:R0}.
 From this Figure, it is clear that the average density, velocity and
temperature (obtained from the numerical simulation) agree very well with the
analytical model, except close to the centre of the cluster. The deviations
in the cluster centre are a direct result of the small number of stars
that is present within the inner $\sim 3$~pc of the cluster (see
Figure \ref{fig:PosR0}).

\begin{figure*}
\centering
\includegraphics{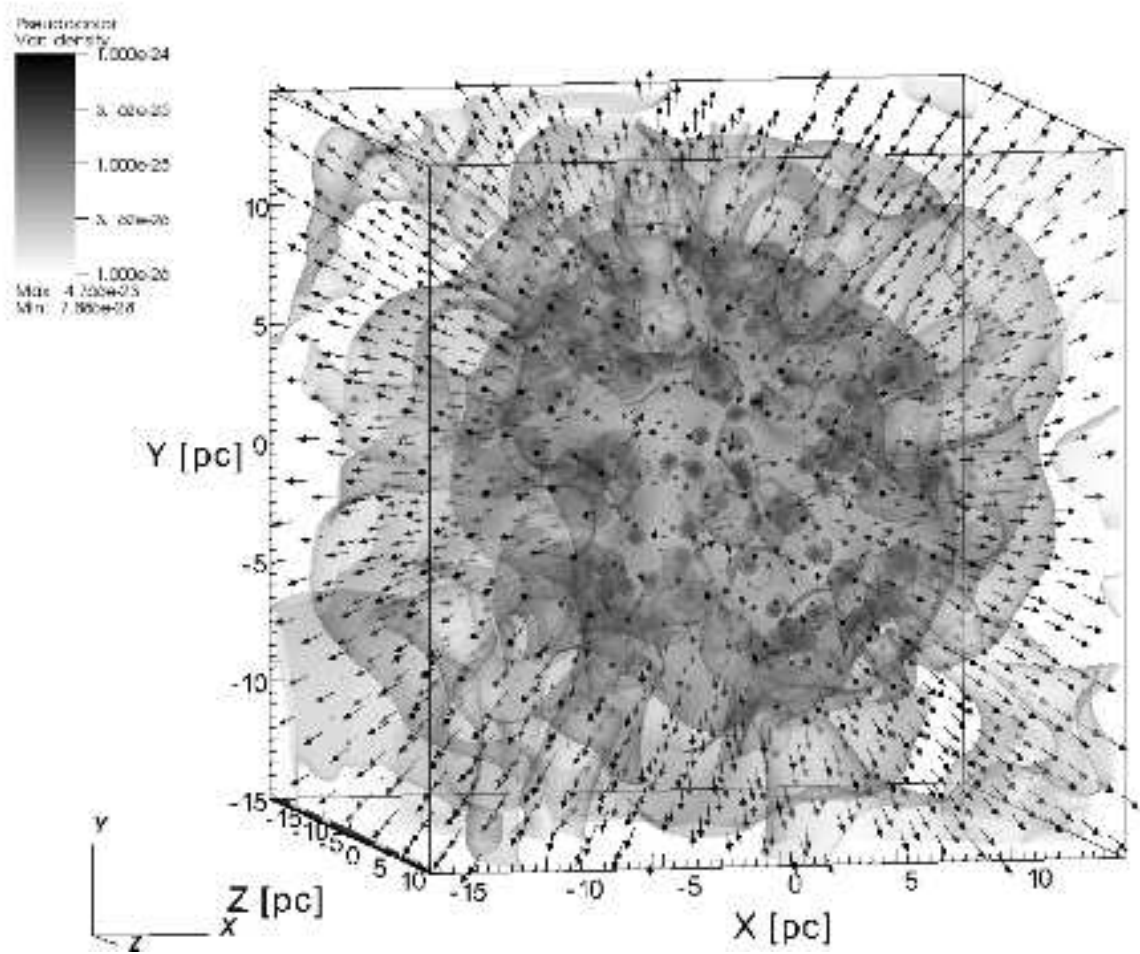}
\caption{3D rendition of the model with $100$ stars distributed
homogeneously (${\alpha = 0}$) inside of a sphere of ${10~\rm{pc}}$ in
radius. In logarithmic grey-scale (colour-scale in the online version)
we present $5$ isosurfaces of density. Depicted by arrows we overlaid
the velocity field, the largest arrow corresponds to a magnitude of
${10^3~\rm{km~s^{-1}}}$ .}\label{fig:DistR0}
\end{figure*}

Figure~\ref{fig:DistR0} illustrates the real complexity of the flow that is obtained
from the numerical simulations. This Figure shows a 3D rendition of
the 10 pc radius star cluster, with 100 stars and $\alpha=0$.
The flow has a sponge-like morphology, with low density
stellar wind cavities immersed in the denser `cluster wind' flow
(composed of shocked stellar wind material). The relatively monotonic
average flow (shown in Figure \ref{fig:R0}) is obtained by averaging
over this complex flow structure.

\begin{figure}
\centering
\includegraphics[width=65mm]{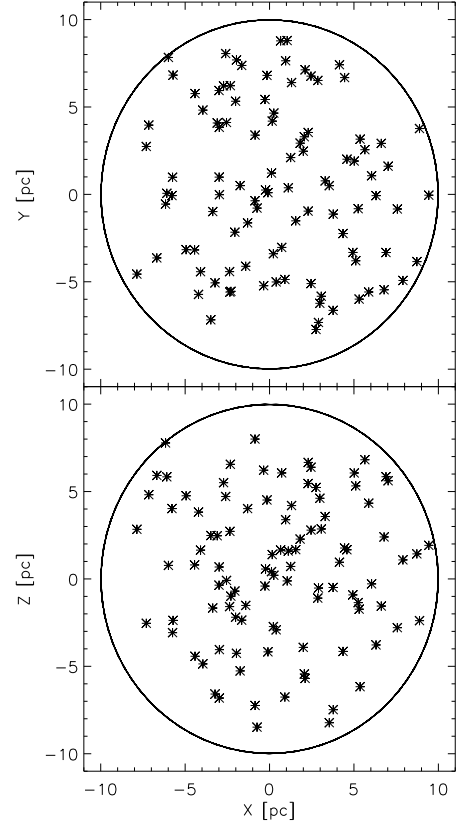}
\caption{Same as Figure~\ref{fig:PosR0}, but
for the $\alpha=-0.5$ model.}\label{fig:PosR05}
\end{figure}

\begin{figure}
\centering
\includegraphics[width=80mm]{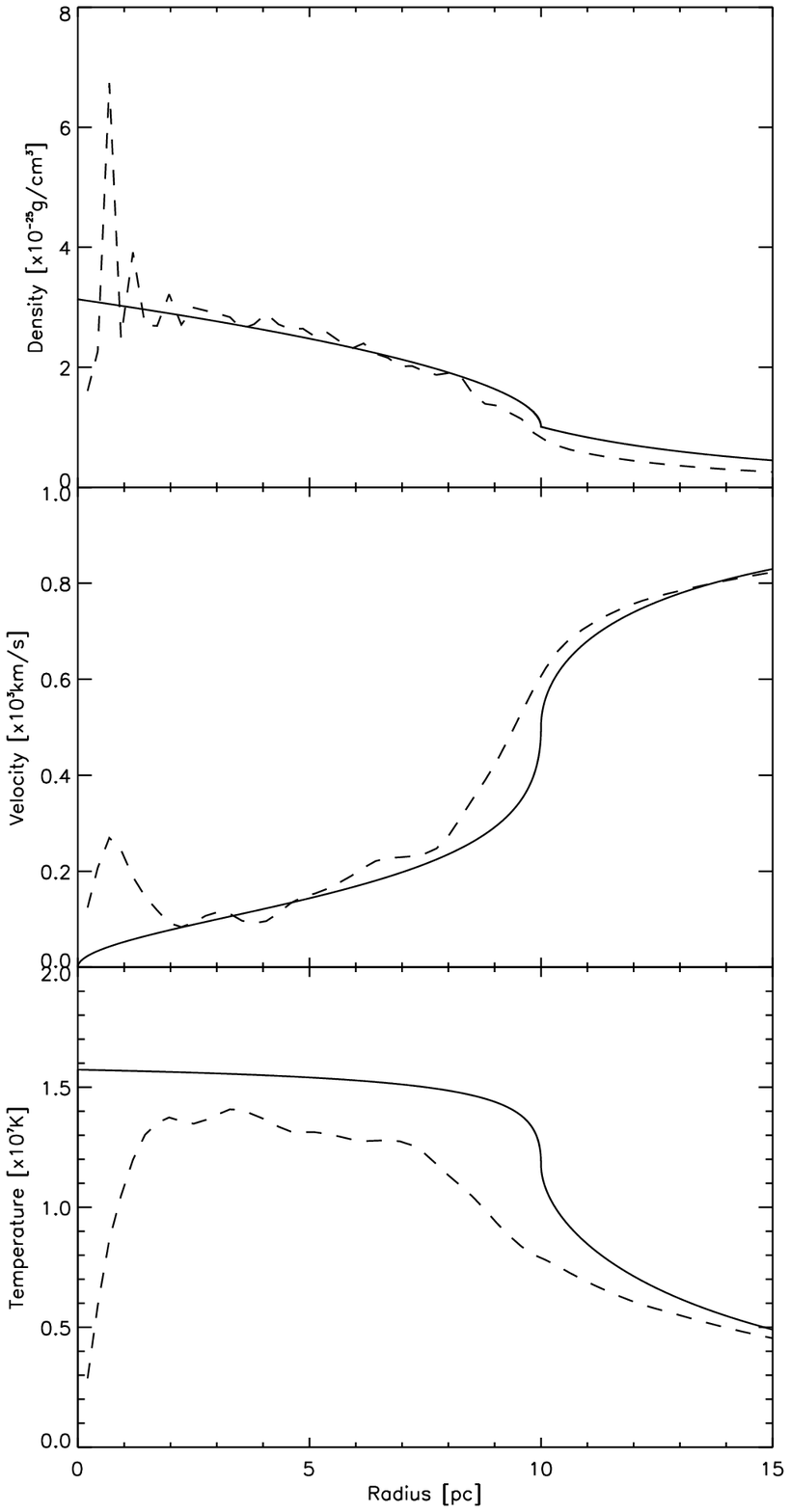}
\caption{Same as Figure~\ref{fig:R0}, but
for the $\alpha=-0.5$ model.}
\label{fig:R05}
\end{figure}

The stellar distribution used for the $\alpha=-0.5$ model
(obtained by sampling the $\alpha=-0.5$ distribution function,
see equation~\ref{eq:Frac} and section~\ref{sec:StarPos}) is shown in Figure~\ref{fig:PosR05}.
The average flow variables as a function of spherical radius
(obtained by averaging over spherical surfaces, see above) are compared
with the solution obtained from the analytic model in Figure \ref{fig:R05}.
Again, a reasonably good agreement is obtained between the analytic
model and the average flow variables computed from the flow
that results from the numerical simulation, except for the inner region
of the cluster.

\subsubsection{The $\alpha_{min}$ $<$ $\alpha$ $<$
$\alpha_c$ case}

%\begin{figure}
%\centering
%\includegraphics[width=80mm ]{./Figuras/Dist_R2.eps}
%\caption{ for $\alpha=-2$ model.}\label{fig:DistR2}
%\end{figure}

\begin{figure}
\centering
\includegraphics[width=65mm]{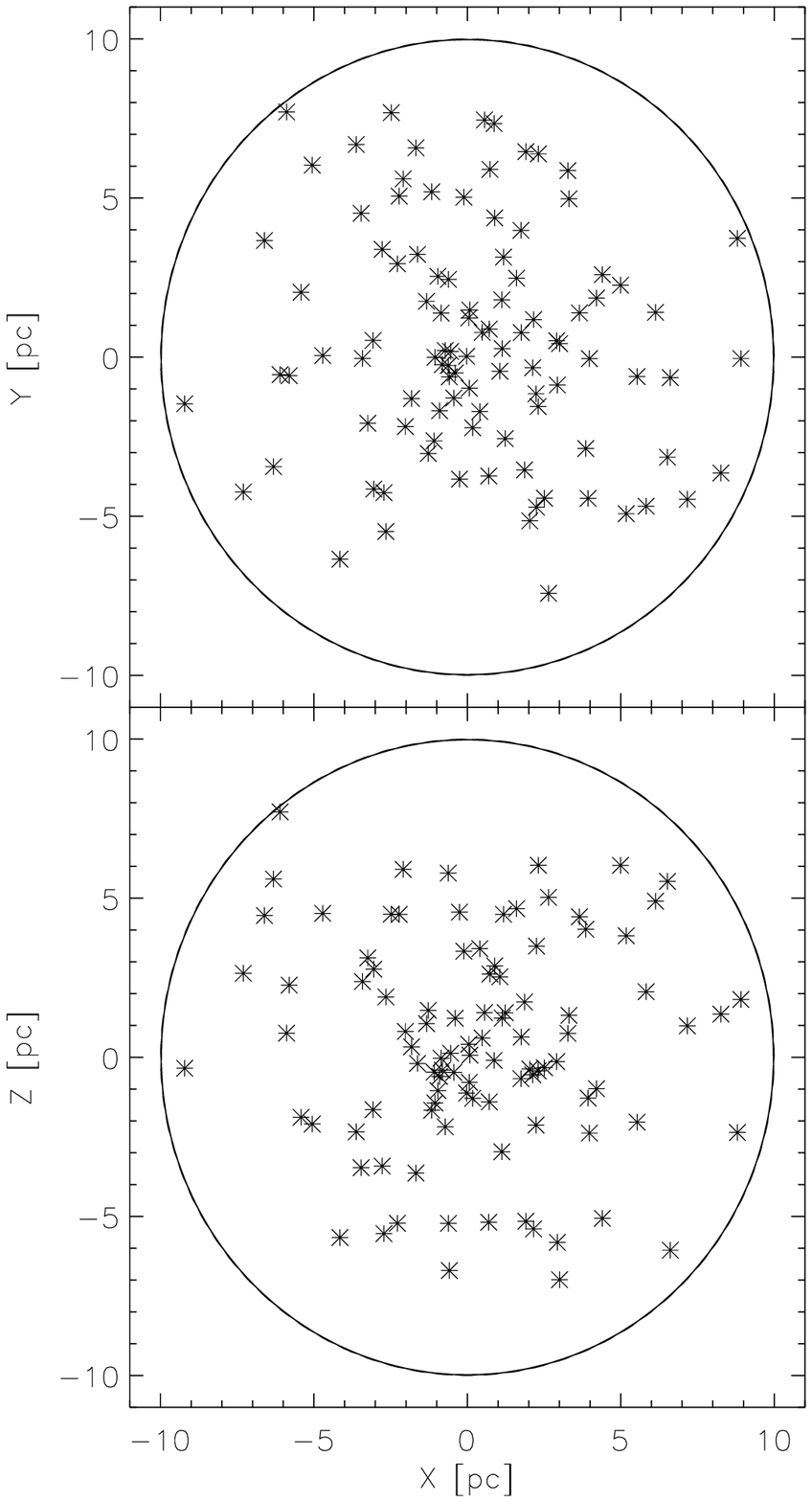}
\caption{Same as Figure~\ref{fig:PosR0}, but for
the $\alpha=-2$ model.}\label{fig:PosR2}
\end{figure}

Figure~\ref{fig:PosR2} shows the stellar positions that
results from sampling the $\alpha=-2$ distribution function.
The stars are highly concentrated around the centre of the cluster.

\begin{figure}
\centering
\includegraphics[width=90mm]{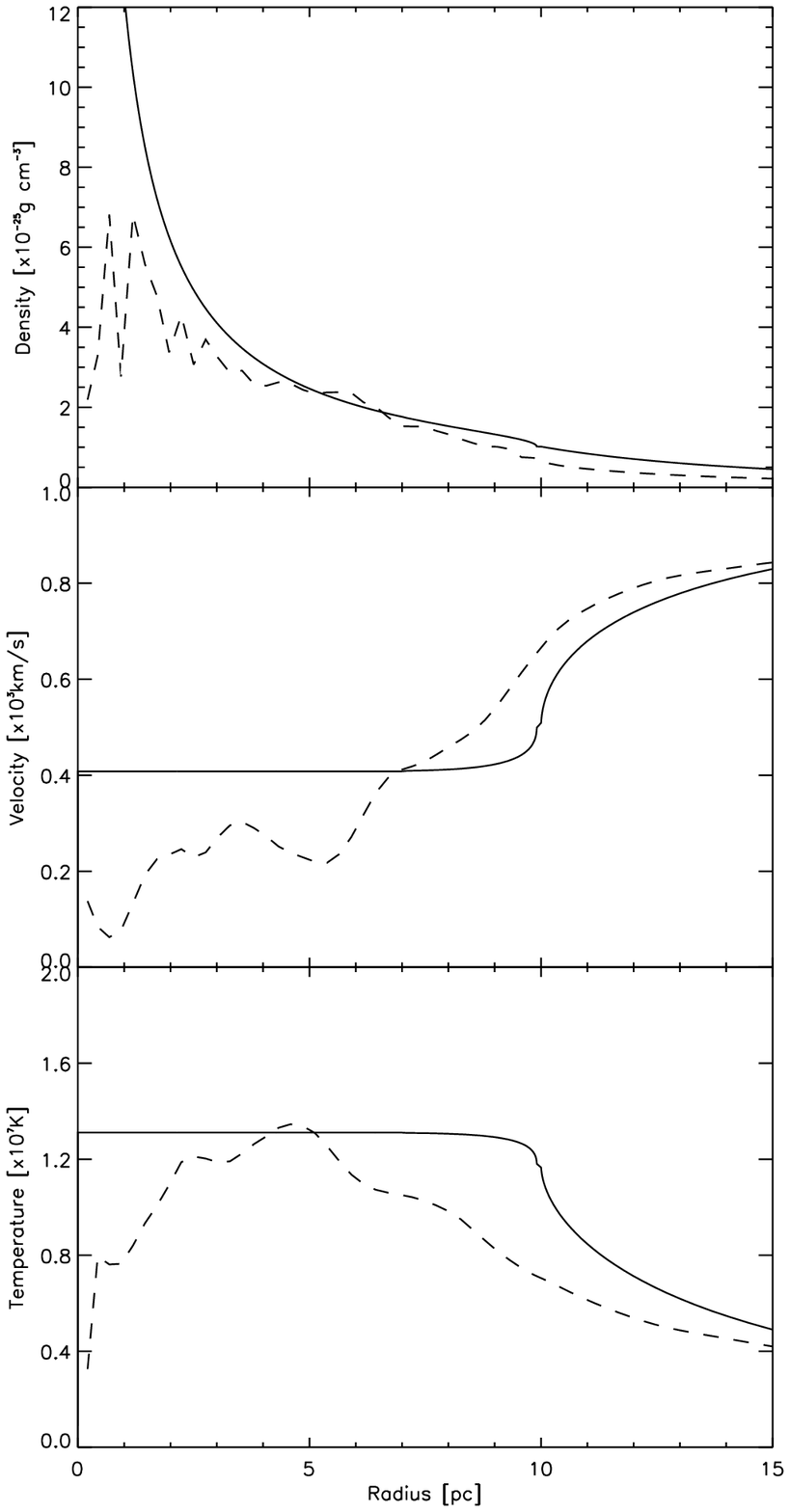}
\caption{Same as Figure~\ref{fig:R0} but for the
$\alpha=-2$ model.}\label{fig:R2}
\end{figure}

Figure~\ref{fig:R2} shows the angularly averaged
velocity, density and temperature
as a function of the radial coordinate. 
The analytical solution has a non-zero, subsonic velocity
and an infinite density
at the central position of the cluster.

Substantial differences
between the numerical and analytical solutions are found for radii
smaller than $\sim 4$~pc. This is a direct result of the undersampling
of the stellar distribution function which occurs as a result of
the ``proximity criterion'' (described in \ref{sec:StarPos}) applied for
placing the stars in the computational grid. For larger radii, a reasonable
agreement between the analytical and numerical results is obtained.\\

\subsubsection{The  $\alpha_{cr}$ $<$ $\alpha$ $<$ $\alpha_{min}$
case}

\begin{figure}
\centering
\includegraphics[width=65mm]{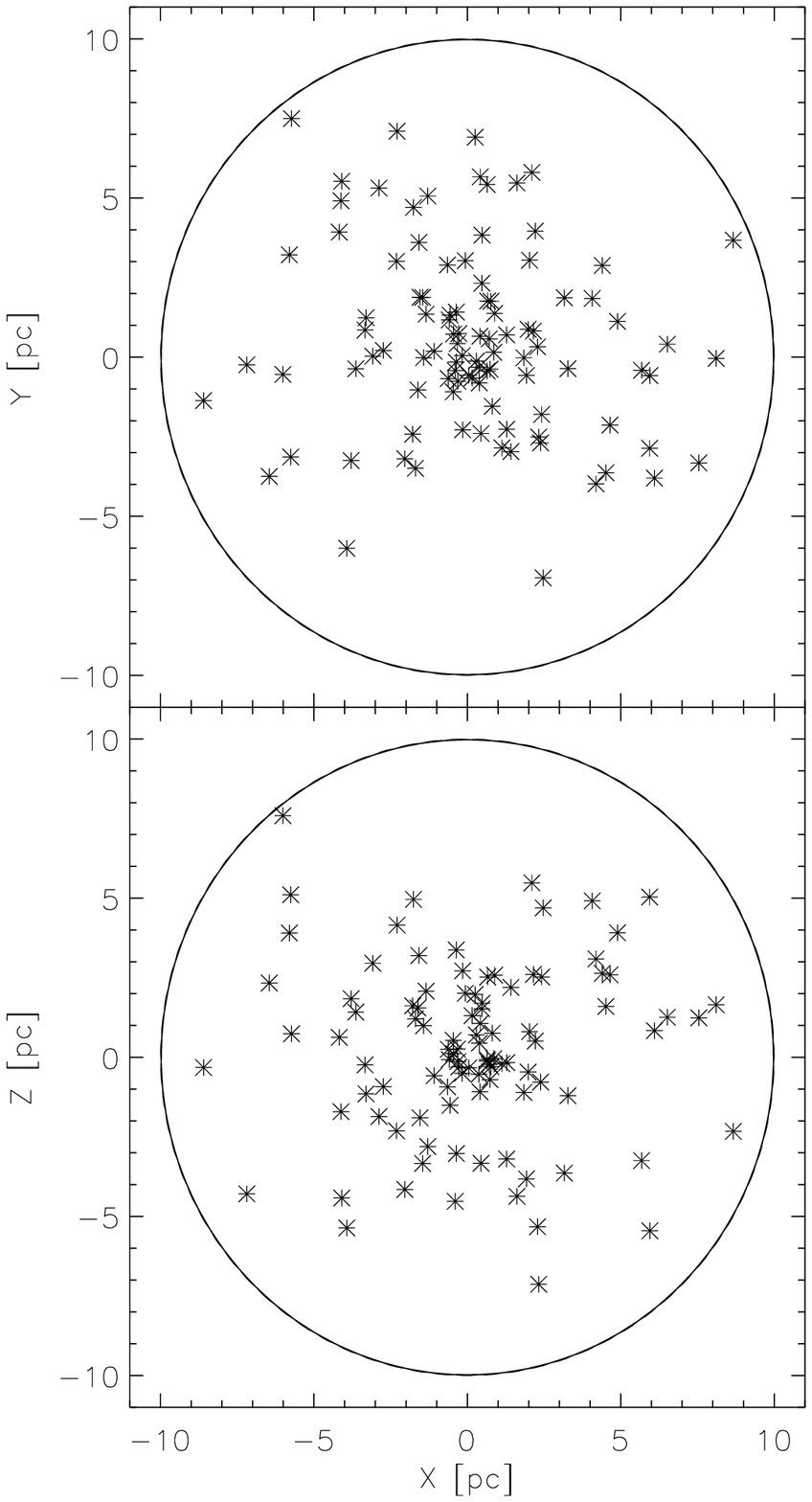}
\caption{Same as Figure~\ref{fig:PosR0} for $\alpha=-2.5$ model.}
\label{fig:PosR25}
\end{figure}

\begin{figure}
\centering
\includegraphics[width=80mm]{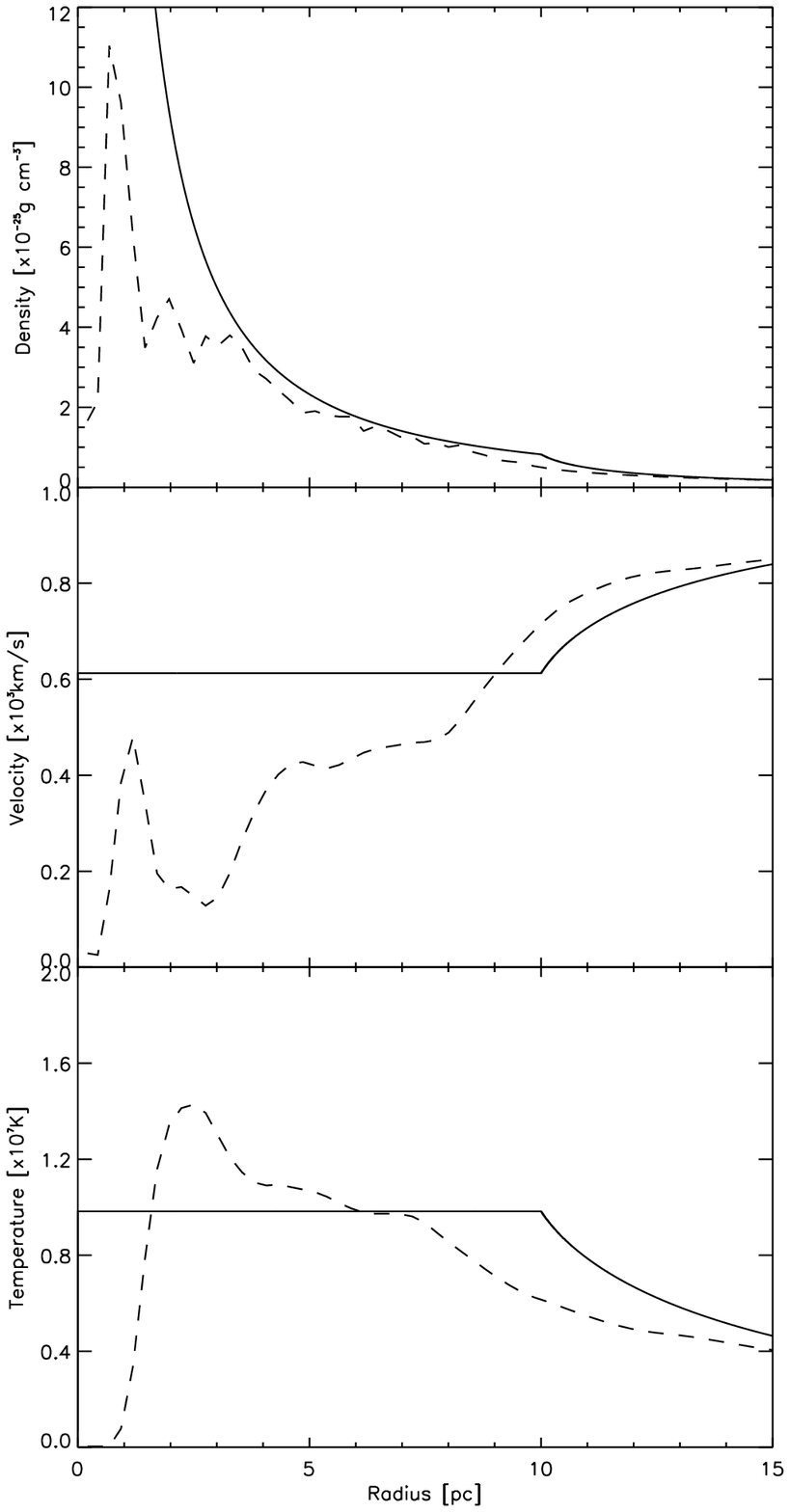}
\caption{Same as Figure~\ref{fig:R0} for $\alpha=-2.5$ model.}
\label{fig:R25}
\end{figure}

Figure~\ref{fig:PosR25} shows the stellar positions that
results from sampling the $\alpha=-2.5$ distribution function.
The analytical solution and the angularly average flow variables
(obtained from the numerical simulation) are shown in Figure~\ref{fig:R25}.
For this model, the analytic solution has a supersonic, outwards velocity
in the inner region of the cluster. Again we obtain substantial deviations
between the analytic and numerical solutions in the central region
of the cluster, and better agreement for larger radii.

\section{Predicted X-ray emission}

We have taken the density and temperature stratifications for the
steady cluster wind flow configurations, and used them to compute the
X-ray emission. We have done this by computing the emission coefficient
in the $0.3\to 2$~ keV photon energy range using the
CHIANTI\footnote{The CHIANTI database and associated IDL procedures, now distributed as version 3.0, are freely available at the following addresses on the World Wide Web: http://wwwsolar.nrl.navy.mil/chianti.html, http://www.arcetri.astro.it/science/chianti/chianti.html, and http://www.damtp.cam.ac.uk/user/astro/chianti/chianti.html} atomic data base and software (see Dere et al.
2001 and references therein). For this calculation, it is assumed
that the ionisation state of the gas corresponds to coronal ionisation
equilibrium, and that the emission is in the low density regime
(i.~e., that the emission coefficient is proportional to the square
of the density).

\begin{figure*}
\centering
\includegraphics[width=100mm]{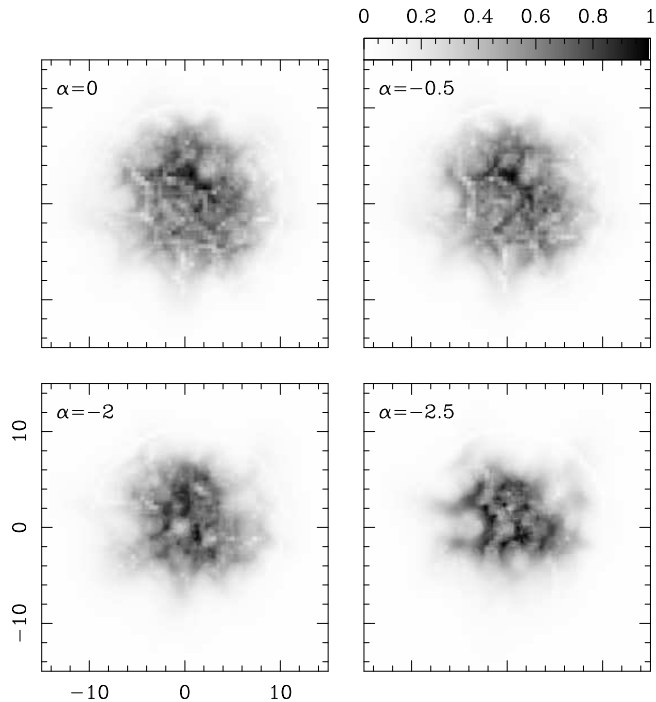}
\caption{X-ray images obtainded by integrating the emission coefficient along
  lines of sight (parallel to $z$-axis) and over energies ranging
  from 0.3 to 2 keV, for the four models described in the text. The images are
  depicted with gray scale (with the fluxes normalized to the maximum value)
  given by the bar to the top of the right-top plot). The scales along the
  $x$- and $y$-axes are given in units of parsecs, with the origin coinciding
  with the barycenter of the stellar distribution of the cluster.}
\label{fig:mapas}
\end{figure*}

We then integrate this emission coefficient along lines of sight, which
are assumed to be parallel to the $y$-axis of the computational grid.
The X-ray maps computed in this way for the models with
$\alpha=0$, $-0.5$, $-2$ and $-2.5$ are shown in Fig. 10.

From these maps, it is clear that the X-ray emission is highly
structured in all models. Actually, some of the structures are seen
in all four models, as the stellar position have been chosen with
the same set of random numbers (but not the same radial positions, see
the discussion at the end of \S 3.2).

\begin{figure}
\centering
\includegraphics[width=100mm]{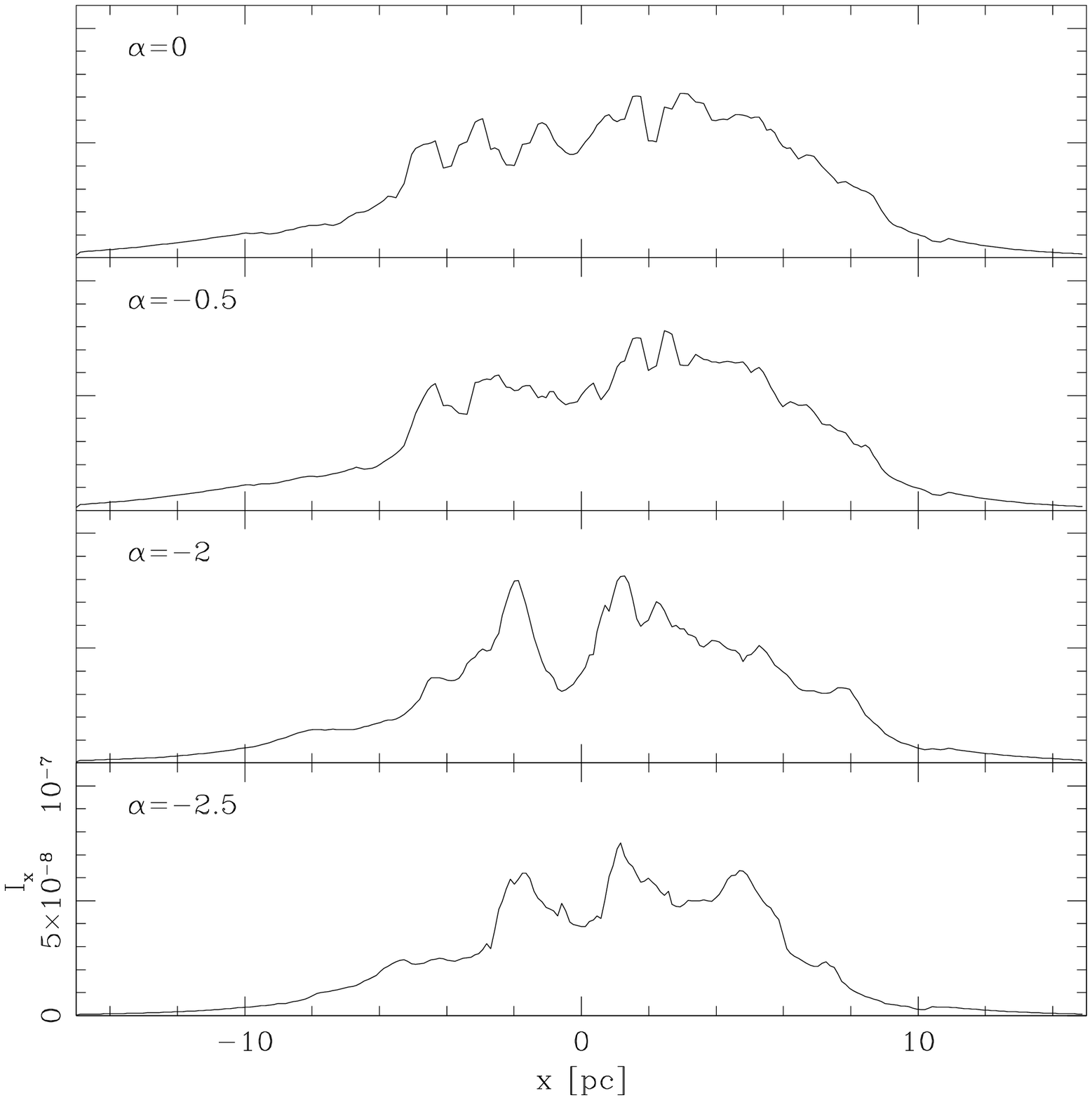}
\caption{X-rays $z=0$ emission profiles for a line of sight at $x$, $y$ =0 for the four x-rays maps in Figure~10.}
\label{fig:int}
\end{figure}

In Figure 11, we show the emission along a line of sight placed at $x$, $y$=0 four X-ray maps of
Figure 10. These cuts show that the $\alpha=0$ and $\alpha=-0.5$
models have a structured but more or less constant emission
within a radius of $\sim 6$~pc (i.~e., the inner 60~\% of
the cluster radius), with wings of lower emission extending
to beyond the outer cluster boundary. The models with
steeper stellar distribution
functions (i.~e., with more negative values of $\alpha$) show
more compact emission structures.

Finally, by integrating the X-ray emission over the whole
emitting volume, we compute luminosities of 0.32, 0.32, 0.26
and 0.20 L$_\odot$ for the $\alpha=0$, $-0.5$, $-2$ and $-2.5$,
respectively. Therefore, we find that the while the predicted
maps differ quite appreciably between the different models,
the X-ray luminosity does not depend strongly on the value
of $\alpha$.

We should note that because the emission coefficient is proportional to
the square of the density, one can use equation (21) to deduce the
scaling of the X-ray luminosity to other model parameters. For
example, a supercluster with $\sim 20000$ stars would have X-ray
luminosities of $\sim 8\to 12\times 10^3$~L$_\odot$ (depending
on the value of $\alpha$, see above).

\section{Conclusions}

In this paper we have extended the analytic cluster wind
model of Paper I to the case of a non-uniform stellar
distribution. In particular, we have studied the case
of a radially dependent, $n(R)=k_c R^\alpha$ power-law
distribution.

Of particular interest is the $\alpha=-2$ distribution, which
corresponds to the stratification of a singular, isothermal,
self-gravitating sphere. Such a structure is of interest for
modelling the wind from a gravitationally bound stellar cluster.
Power-law stellar distributions with other values of $\alpha$ do not
have a clear physical justification, but can be considered as a
parametrization of stellar distributions with different degrees
of central condensation.

We find that for shallow distributions, with $-1<\alpha < 0$,
the cluster wind has zero velocity in the cluster centre. However,
for more negative $\alpha$ values the cluster wind has a non-zero,
outwards directed velocity (subsonic for $\alpha_{min}<\alpha < -1$ and
supersonic for $-3<\alpha<\alpha_{min}$) in the centre of the cluster.
The solutions with a non-zero central velocity have an infinite
central density for the cluster wind.

We have then compared the analytic cluster wind solutions with
3D numerical simulations. For carrying out the simulations, we
consider the winds from 100 stars, with a spatial distribution
obtained by statistically sampling the appropriate stellar distribution
function. We then carry out angular averages of the computed
flow variables, and compare the radial dependence of these
averages with the predictions obtained from the analytic model.

For different values of $\alpha$, we obtain a good agreement
between the analytic and numerical predictions in the outer
regions of the cluster. However, the analytic and numerical
solutions have large differences in the central region of the
cluster. These differences are a direct result of the fact
that only a small number of stars are present in this spatially
reduced region, and therefore the continuous mass and energy
source distribution assumed in the analytic model is inappropriate
for describing the real cluster wind flow.

Finally, we have obtained predictions of the X-ray emission from
our simulated cluster wind flows. Our models have quite low
$\sim 0.2\to 0.3$~L$_\odot$ X-ray luminosities, as a result of
the fact that only 100 stars were included (due to the constraints
imposed by the numerical resolution of the simulations). However,
the scaling laws of the analytic model imply that models with
20000 stars (i.~e., with the number of stars of a large supercluster)
will have X-ray luminosities of $\sim 10^4$~L$_\odot$ (see section 4).
At distances of $\sim 4$~Mpc (i.~e., the distance to M~82), this
would produce an X-ray flux of
$\approx 2\times 10^{-14}$ erg~s$^{-1}$~cm$^{-2}$. Such a flux is only
2~\% of the X-ray flux observed for the Arches cluster close
to the galactic centre (see Yusef-Zadeh et al. 2002), but might
be within range for possible future observations.

An important remaining problem is that H$\alpha$ emission has been
observed from many superclusters. Our models have gas temperatures
in excess of $10^7$~K, and therefore do not produce such an emission.
Therefore, the observed H$\alpha$ emission must be coming from
another component, which could be the remnants of dense interstellar
clouds which were present in the region when the cluster was formed.
The presence of such dense, lower temperature structures within
the cluster wind flow could be explored in the future with
numerical simulations similar to the ones which we have shown
in the present paper, but that need to include the radiative cooling.

\section*{acknowledgements}
This work was supported by the CONACyT
grant 46828-F, the DGAPA (UNAM) grant IN~108207
and the ``Macroproyecto
de Tecnolog\'\i as para la Universidad de la Informaci\'on y la
Computaci\'on'' (Secretar\'\i a de Desarrollo Institucional de la UNAM,
Programa Transdisciplinario en Investigaci\'on y Desarrollo
para Facultades y Escuelas, Unidad de Apoyo a la Investigaci\'on en
Facultades y Escuelas).

We thank Enrique Palacios and Mart\'\i n Cruz for supporting the
servers in which the calculations of this paper were carried out.

\bsp
\label{lastpage}
\end{document}